\def\year{2018}\relax
\documentclass[letterpaper]{article} 
\usepackage{aaai18}  
\usepackage{times}  
\usepackage{helvet}  
\usepackage{courier}  
\usepackage{url}  
\usepackage{graphicx}  
\usepackage{amsmath}
\usepackage{bm}
\usepackage{amsfonts}
\usepackage{color}
\usepackage{subcaption}
\usepackage{booktabs}
\usepackage{algorithm}
\usepackage[noend]{algpseudocode}
\usepackage[subtle,mathspacing=normal,floats=normal]{savetrees}

\newcommand*\samethanks[1][\value{footnote}]{\footnotemark[#1]}

\begin{document}
%
\title{Hawkes Processes for Invasive Species Modeling and Management}
\author{Amrita Gupta\thanks{Corresponding author.}, Mehrdad Farajtabar, Bistra Dilkina\samethanks and Hongyuan Zha\\
School of Computational Science and Engineering\\
Georgia Institute of Technology\\
Atlanta, Georgia 30332\\
\texttt{agupta375@gatech.edu, mehrdad@gatech.edu, bdilkina@cc.gatech.edu, zha@cc.gatech.edu}
}
\maketitle
\begin{abstract}
The spread of invasive species to new areas threatens the stability of ecosystems and causes major economic losses in agriculture and forestry. We propose a novel approach to minimizing the spread of an invasive species given a limited intervention budget. We first model invasive species propagation using Hawkes processes, and then derive closed-form expressions for characterizing the effect of an intervention action on the invasion process. We use this to obtain an optimal intervention plan based on an integer programming formulation, and compare the optimal plan against several ecologically-motivated heuristic strategies used in practice. We present an empirical study of two variants of the invasive control problem: minimizing the final rate of invasions, and minimizing the number of invasions at the end of a given time horizon. Our results show that the optimized intervention achieves nearly the same level of control that would be attained by completely eradicating the species, with a 20\% cost saving. Additionally, we design a heuristic intervention strategy based on a combination of the density and life stage of the invasive individuals, and find that it comes surprisingly close to the optimized strategy, suggesting that this could serve as a good rule of thumb in invasive species management.
\end{abstract}

\section{Introduction}
\label{sec:intro}
Network diffusion models are a powerful tool for studying dynamic processes like the spread of influence and information through social networks \cite{kempe2003maximizing,yang2010modeling,romero2011differences,farajtabar2016multistage}, the dispersal of species through a landscape \cite{sheldon2010maximizing}, disease contagion in populations \cite{eames2002modeling}, and signal transduction in cell signaling networks \cite{nalluri2017determining}. The ability to model the dynamics of these diffusion processes enables the development of strategies for steering them towards desirable outcomes. For instance, in conservation planning, one might selectively add land parcels to an existing protected area to facilitate the colonization of new habitat by a certain species \cite{sheldon2010maximizing}. In order to contain a disease or contamination, one can strategically block transmission along a set of links \cite{kimura2008minimizing,khalil2014scalable}.

Two of the most studied network diffusion models are the independent cascade (IC) model and the linear threshold (LT) model \cite{kempe2003maximizing}. In both, the spreading process is modeled as an activation of nodes over discrete time steps. Each node in the network is in a binary state (active or not), and nodes are activated by their active neighbors. In both the IC and LT models, once a node is active it remains so for the rest of the diffusion process, an assumption that is appropriate for modeling the spread of irreversible phenomena, e.g. the adoption of a product, infection by a disease that confers permanent immunity, or propagation of invasive species.

However, many network diffusion processes exhibit \emph{non-progressive} cascades where an active node can become inactive probabilistically at each time step, so that the state of a node fluctuates over time. For example, in species dispersal, a previously occupied habitat patch may become unoccupied \cite{sheldon2010maximizing}, or in the spread of a flu-like illness, a patient may recover but be susceptible to reinfection. In this setting, repeated exposure to activation events plays an important role in continuing the diffusion process by reactivating nodes that have become inactive. Sometimes, exposure to multiple activations can also cause a node to become ``more'' active, e.g. the posting frequency of an individual social media user can increase due to high activity in their network. In these cases, it is more fitting to model the state of a node as a time-varying, real- or continuous-valued function as opposed to binary states. Furthermore, activation events typically arrive continuously rather than in discrete time steps, warranting the diffusion process to be modeled in continuous time.

\emph{Temporal point processes} offer a framework for modeling diffusion processes with both continuous activity states and continuous time. The activity of a node can be characterized by a parameter $\lambda$ representing the rate at which the node stochastically tries to generate events. This $\lambda$ parameter itself can be responsive to activations arriving at the node, thereby capturing self-exciting behavior in the diffusion process. Temporal point processes have recently been applied to modeling several diffusion processes like the activity of Twitter users \cite{farajtabar2014shaping}, criminal activity \cite{mohler2011self}, and the spread of avian flu \cite{kim2011spatio}. Similar to our application, \cite{balderama2012application} use a spatiotemporal point process model to characterize the spread of an invasive banana plant, although they do not consider any control mechanisms.

In terms of controlling diffusion processes, a variety of intervention actions have been analyzed in the discrete-time, binary-state setting, such as selecting source nodes for initiating cascades \cite{kempe2003maximizing} and modifying network connectivity to guide the diffusion by adding or removing nodes \cite{sheldon2010maximizing} and edges \cite{kimura2008minimizing,khalil2014scalable} or modifying edge weights \cite{wu2015efficient}. In contrast, there has been relatively little work on controlling dynamics in network temporal point processes. One possible control action is to manipulate the activity rate parameters $\lambda$ at specific nodes, e.g. by incentivizing social media users to post more frequently. Steering user activity in this manner was first considered in~\cite{farajtabar2014shaping}, and \cite{farajtabar2016multistage} used the same intervention to develop a multistage strategy for shaping network diffusion with applications to mitigating fake news \cite{farajtabar2017fake}. Recent work has also applied methods from stochastic differential equations to find the best intensity for information guiding~\cite{wang2016steering} and achieving highest visibility~\cite{zarezade2017redqueen}. In our work, a \emph{discrete} intervention for network point processes is considered for the first time that, unlike the above, modifies the activity rate parameter at select nodes by \emph{deleting the history} of the point process.



Our work is motivated by the invasive species management problem in biodiversity conservation. The spread of non-native species to new areas is a cause of major concern, because they harm native species through predation, competition, disease or by otherwise disrupting food webs and ecosystem processes. These adverse effects have generated significant interest in limiting their spread. In particular, it is often important to eradicate invasive species to prevent irreversible change to ecosystems, but their removal can be prohibitively costly. In light of this, a common objective is to optimize the location of control efforts in order to maximize the efficacy of the intervention. We derive a novel approach for finding an optimal set of locations at which to remove individuals of an invasive species given a fixed budget. Although our work is motivated by a specific problem in environmental sustainability, the novel computational problem it poses appears in other domains that can be modeled using temporal point processes, such as mitigating the spread of pandemic infections using vaccination programs. The computational approach we develop here can be generalized to these broader applications.


\section{Invasive Species Management and Hawkes Processes}
\label{sec:invasive-management-hawkes-process}
\subsection{Problem Statement}
\label{subsec:problem-statement}

In the invasive species management problem, the goal is to identify locations at which to eradicate invasive individuals in order to minimize the spread of the species through the landscape. Let $L$ be a set of distinct land parcels corresponding to basic units of management. An invasive species is observed to be proliferating and dispersing through the landscape until a given time $\tau$, when an intervention is performed by eliminating all invasive individuals present before $\tau$ in a set of land units $U\subseteq L$. Each land unit $i\in L$ has an associated cost $c_i$ reflecting economic land management costs or effort needed to eradicate the invasive individuals, and the total cost of the intervention cannot exceed a given budget $\mathcal{B}$. A feasible intervention plan is therefore a set of land parcels $U$ with total intervention cost within $\mathcal{B}$. After the intervention, the invasive species continues to spread until time $T>\tau$, but without the proliferative influence of the individuals eradicated at time $\tau$. Our goal is to find a feasible intervention plan that minimizes the degree to which the landscape is affected by the invasion.

To formulate the invasive species management problem as a network diffusion optimization problem, we consider a landscape $L$ consisting of $n$ distinct land parcels modeled as nodes $V$ in a graph, with edges between nodes that are close enough for dispersal to occur. The appearance of new invasive individuals in the network is modeled as a multivariate Hawkes process (see Section \ref{subsec:hawkes-background} for a more rigorous treatment), where an invasion event at node $i$ at time $s$ is denoted $(i,s)$. Indexing invasion events by $e$, the history of the network diffusion process up to immediately before some time $t$ is $\mathcal{H}_{t-}:=\left\lbrace (i_e,s_e) | s_e < t\right\rbrace$.

Invasive species can be introduced at any time by carriers like wind, animals or humans. These arrivals are called \emph{exogenous} invasions, and their rate can vary spatially depending on landscape features or human activity. The instantaneous rate at which individuals are introduced to node $i$ at time $t$ is denoted by $\mu_i(t)$, and represents the probability of an exogenous invasion event in a small time window $\left[t, t+dt\right)$. Once an invasive individual has become established, it survives for an average lifetime $\beta$. Since many invasive species mature early and have short life expectancy \cite{sakai2001population}, we assume an individual born at $s_e$ faces a constant risk of death $\omega=\frac{1}{\beta}$, so that the probability of the individual surviving until time $t$ is given by the survival function $e^{-\omega(t-s_e)}$. While the individual survives, it initiates \emph{endogenous} invasions, e.g. by releasing offspring. The likelihood of the offspring of an individual at location $i$ dispersing to location $j$ depends on an edge weight $a_{ij}$ between the two nodes, which can be, e.g., a decaying function of the distance between $i$ and $j$ \cite{arim2006spread}.

All these effects together influence the rate at which new individuals appear in a given node $i$ at time $t$, or the intensity $\lambda_i(t)$. This represents the conditional probability of observing an invasion event in a small time window $\left[t, t+dt\right)$ given the history $\mathcal{H}_{t-}$.
\begin{equation}
\centering
\lambda_i(t) = \mu_i(t) + \sum_{\substack{(j_e,s_e) \in \mathcal{H}_{t-}}} a_{ij_e} \cdot e^{-w(t-s_e)}
\end{equation}
The first term $\mu_i(t)$ is the rate of exogenous invasion events at node $i$, and the summation term captures the contribution of past invasion events $(j_e,s_e)$ in the network towards endogenous invasions in node $i$ at time $t$.

\subsubsection{Control Objectives}
Given the graph representing our landscape and the invasion process dynamics described above, we can quantify the degree to which the landscape is affected by the invasive species spread at the end of our planning horizon $T$ in a number of ways. One reasonable goal is to minimize the rate of invasions at time $T$, captured by $\lambda(t;u)$. Since $\lambda(t;u)$ depends on events that will stochastically occur between $\tau$ and $t$, it will vary across different realizations of the stochastic process, so instead we aim to minimize the total \emph{expected} intensity at time $T$. Let ${\eta_i(t;u) = \mathbb{E}\left[ \lambda_i(t;u) \right]}$, where the expectation is taken over all possible realizations of the stochastic process.
\begin{itemize}
\item[] \textbf{Given:} A graph $G(V,E)$ representing landscape $L$, edge weights $A$ with $a_{ij} = 0$ for $(i,j)\notin E$ and $a_{ij}>0$ for $(i,j)\in E$, intervention time $\tau$ and finite time horizon $T$, intervention costs $c_i$ for each node $i\in V$ and budget $\mathcal{B}$.
\item[] \textbf{Find:} A feasible intervention plan consisting of nodes $U\subseteq V$ such that $\sum_{i\in U} c_i \leq \mathcal{B}$, that minimizes $\sum_{i\in V} \eta_i(T;u)$.
\end{itemize}

Another plausible goal is to minimize the \emph{total expected number} of invasions that occur until time $T$, since the ecological damage resulting from invasions is often a function of the population size \cite{blackwood2010cost}. We cannot affect the process until $\tau$, so this amounts to minimizing the number of invasions in the interval $\left[\tau, T\right)$. We store the number of invasion events at each node over time using an $n$-dimensional counting process where $\mathcal{N}_i(t;u)$ represents the number of invasive species individuals that have appeared in cell $i$ by time $t$. Then, given the same inputs as before,
\begin{itemize}
\item[] \textbf{Find:} A feasible intervention plan consisting of nodes $U\subseteq V$ such that $\sum_{i\in U} c_i \leq \mathcal{B}$, that minimizes $\sum_{i\in V} \mathbb{E}\left[\mathcal{N}_i(T)\right]$.
\end{itemize}

\subsection{Hawkes Processes}
\label{subsec:hawkes-background}
A multivariate Hawkes process can be thought of as a spatiotemporal point process, which is a random collection of points representing the time and location of events. An $n$-dimensional point process can be described by a counting process $\mathcal{N}(t) = \left(\mathcal{N}_1(t),\cdots, \mathcal{N}_n(t)\right)^\top$ where $\mathcal{N}_i(t)$ is the number of events occurring at location $i$ before time $t$. The behavior of the process can be characterized by the conditional intensity $\lambda(t)$. Given the history of the process up to time $t$, $\mathcal{H}_{t-}$, the expected number of events in a small time window $\left[t,t+dt\right)$ is given by $\mathbb{E}\left[d\mathcal{N}(t) | \mathcal{H}_{t-}\right] = \lambda(t)dt$.

Hawkes processes model self-exciting phenomena in which the occurrence of events causes additional events to be more likely, such as social media posts spurring reposts \cite{farajtabar2014shaping}, earthquake aftershocks inducing further aftershocks \cite{ogata2006space}, neuronal spike trains causing neighboring neurons to fire \cite{krumin2010correlation}, and in this work, an invasive species individual causing another individual to appear at the same or other nodes. This self-exciting behavior is modeled using a history-dependent intensity of the form:
\begin{align}
\label{eq:intensity-hawkes}
\lambda_i(t) &= \mu_i(t) + \sum_{e: t_e < t} \phi_{i j_e} (t, t_e)\\
&= \mu_i(t) + \sum_{j=1}^n \int_{0}^t\phi_{ij}(t,s) d\mathcal{N}_j(s)
\end{align}
$\phi_{ij}(t,s)$ is called the impact function and captures the temporal influence of an event at location $j$ at time $s$ on the occurrence of events at location $i$ at time $t\geq s$.  Here, the first term $\mu_i(t)$ is the exogenous event intensity, from outside the network and independent of the history, and  the second term $\sum_{e: t_e < t} \phi^{i j_e} (t, t_e)$ is the endogenous event intensity, modeling influence and interaction within the network. Defining $\Phi(t,s)=[\phi_{ij}(t,s)]_{i,j = 1\ldots n}$, $\lambda(t) = (\lambda_1(t) , \ldots, \lambda_n(t))^{\top}$,  and $\mu(t) = (\mu_1(t) , \ldots, \mu_n(t))^{\top}$, we can compactly rewrite Eq~\eqref{eq:intensity-hawkes} in matrix form:
\begin{equation}
\label{eq:intensity-hawkes-compact}
\lambda(t) = \mu(t) + \int_{0}^t \Phi(t,s) d\mathcal{N}(s)
\end{equation}

A common choice of impact function is the truncated exponential function $\Phi(t,s) = A e^{-\omega(t-s)}\cdot \textbf{1}_{\geq0}(t-s)$ where $\phi_{ij}(t,s) = a_{ij}e^{-\omega(t-s)}\cdot \textbf{1}_{\geq0}(t-s)$. The coefficient $a_{ij}$ represents the strength of the influence of $j$ on $i$, and the influence of an event that occurs at time $s$ is 0 before $s$ and decays off after $s$ (e.g. a social media post becomes less relevant, an infected person becomes less contagious, or an invasive species becomes less likely to survive and reproduce). Since the intensity at any time $t'$ only depends on the history of events up to time $t'$, we can also define the \emph{state} at any time $t'$ as $y(t') := \int_{0}^{t'}  e^{-\omega(t'-s)} \, d\mathcal{N}(s)$, capturing the current effect of all events that have happened at each node up to time $t'$. Then,
\begin{align*}
\label{eq:intensity-hawkes-split}
\lambda(t) &= \mu(t) + A y(t) = \mu(t) + \int_{0}^t A e^{-\omega(t-s)} d\mathcal{N}(s)\\
&=\mu(t) + \underbrace{\int_{0}^\tau A e^{-\omega(t-s)} d\mathcal{N}(s)}_{\text{events before $\tau$}} + \underbrace{\int_{\tau}^t A e^{-\omega(t-s)} d\mathcal{N}(s)}_{\text{events after $\tau$}}\\
&=\mu(t) + A e^{-\omega (t-\tau)} y(\tau) + \int_{\tau}^t A e^{-\omega(t-s)} d\mathcal{N}(s)
\end{align*}

\section{Our Approach to Discrete Interventions in Hawkes Processes}

\begin{figure*}[t]
\label{fig:intervention-illustration}
\centering
	\begin{subfigure}[t]{0.19\textwidth}
		\includegraphics[width=\textwidth]{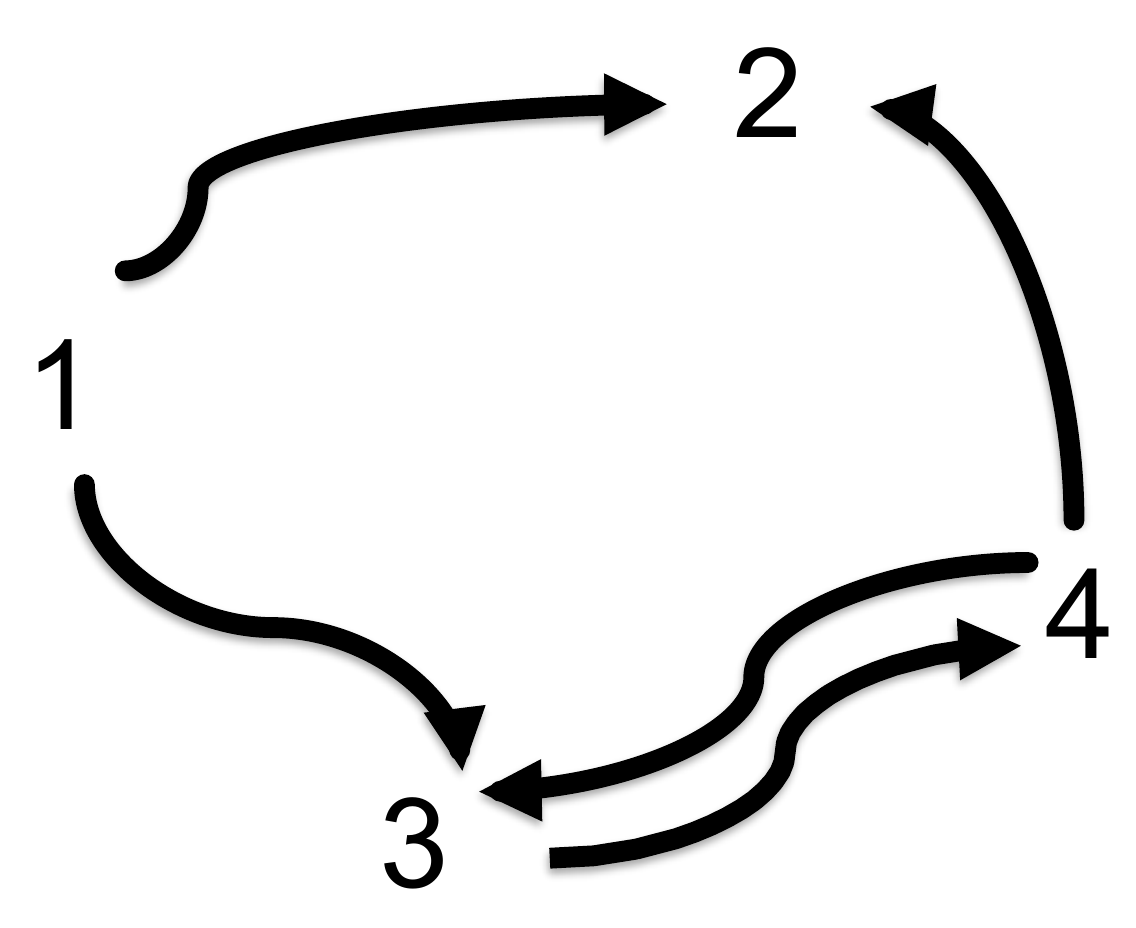}
        \caption{A sample network.}
	\end{subfigure}
    ~
    \begin{subfigure}[t]{0.23\textwidth}
		\includegraphics[width=\textwidth]{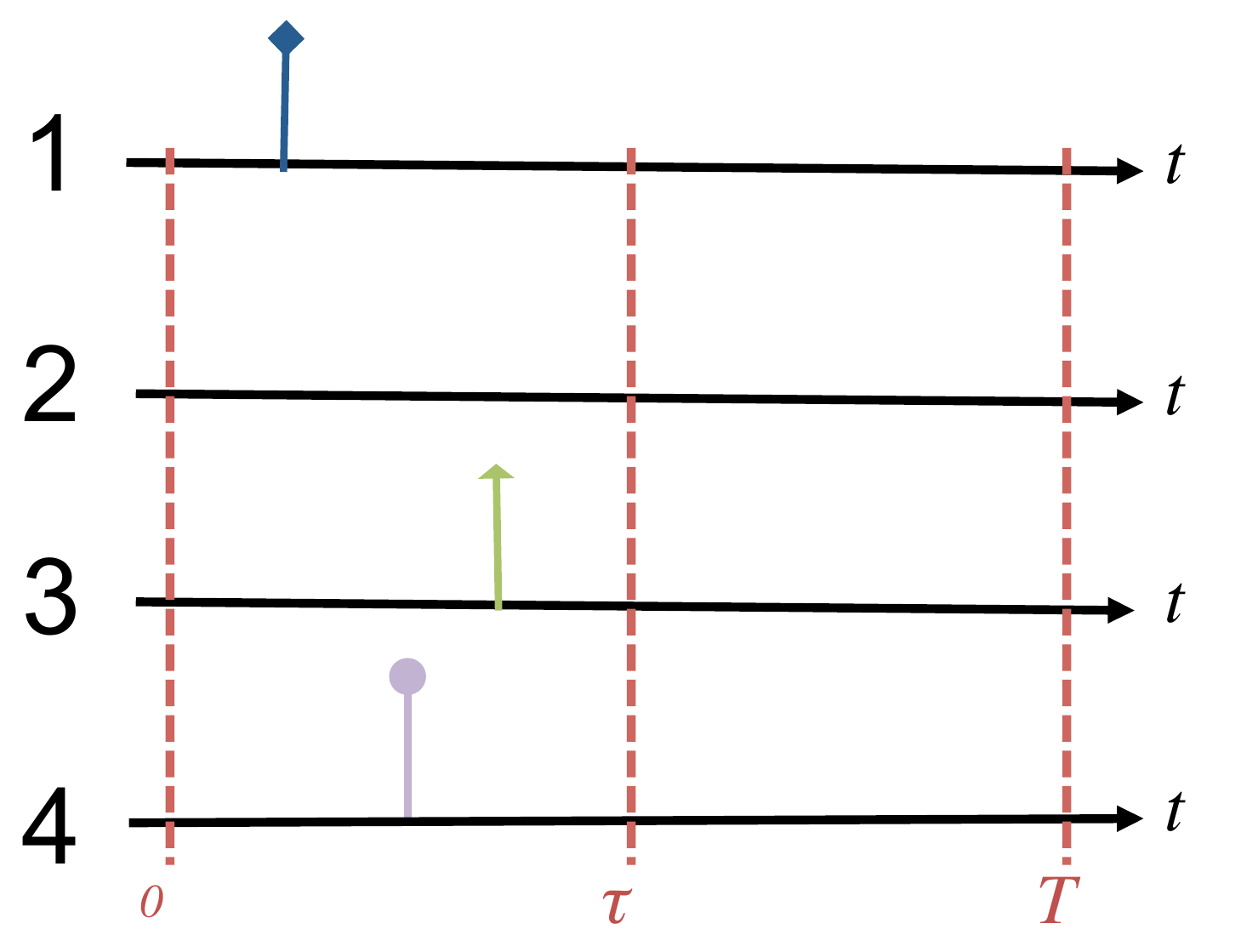}
        \caption{Events before $\tau$.}
	\end{subfigure}
    ~
    \begin{subfigure}[t]{0.23\textwidth}
		\includegraphics[width=\textwidth]{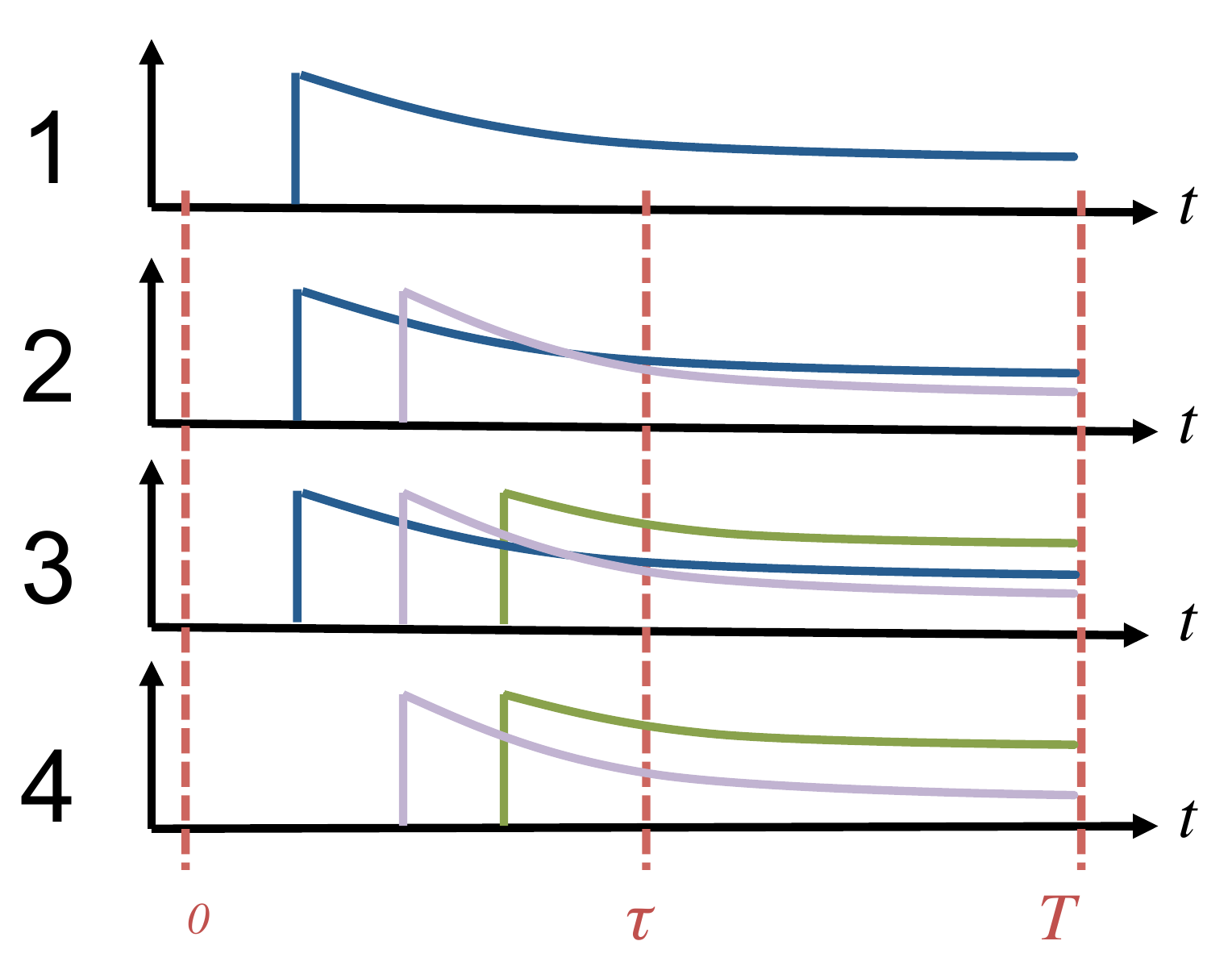}
        \caption{Intensity from events.}
	\end{subfigure}
    ~
    \begin{subfigure}[t]{0.23\textwidth}
		\includegraphics[width=\textwidth]{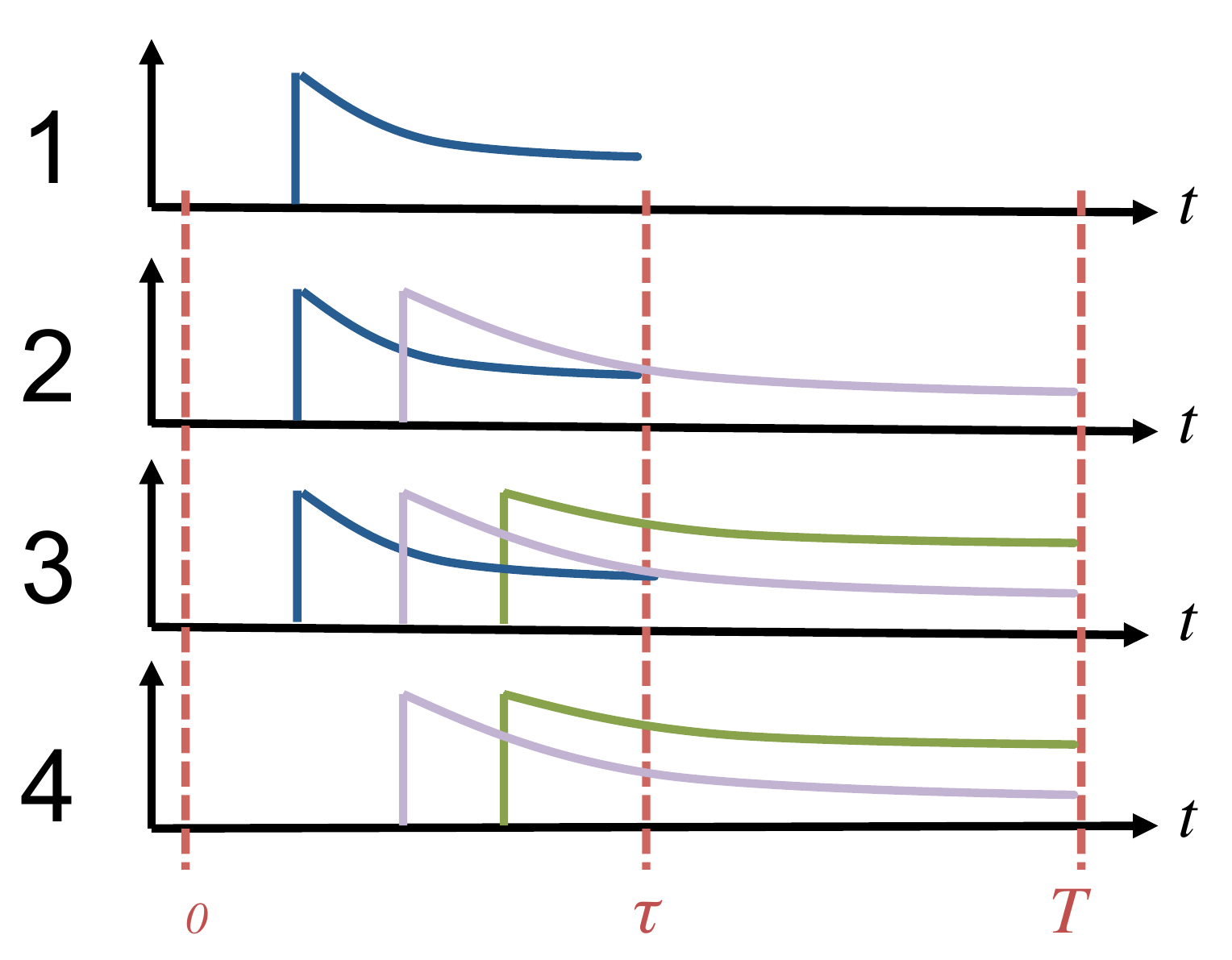}
        \caption{Intensity if event at 1 removed.}
	\end{subfigure}
\caption{A sample network and event history up to $\tau$. Each event contributes to the intensities at the event's node and its neighbors. If the event at node 1 is deleted at $\tau$, its contribution to intensities for $t>\tau$ disappears.}
\end{figure*}

Given an invasion process starting at time $t_0=0$, suppose we plan to perform a management action at time $\tau > t_0$ to steer the invasion process over the landscape network towards some objective at an arbitrary time $T > \tau$. Our management action entails the removal of all invasive individuals at a given set of locations $U$ (see Figure \ref{fig:intervention-illustration}). This can be thought of as deleting events at specific locations from the history of the Hawkes process, or alternatively as resetting the state of those locations to 0 at time $\tau$. Therefore, for $t > \tau$ we have the intervention-dependent intensity:
\begin{equation}
	\begin{split}
	\lambda(t; u) 	&= \mu(t) +  A e^{-\omega (t-\tau)} (u \circ y(\tau)) \\
					&+ \int_{\tau}^{t} Ae^{-\omega(t-s)} \, d\mathcal{N}(s;u)
	\end{split}
\end{equation}
where $\circ$ denotes element-wise product. Vector $u$ encodes our management action (intervention) where $u_i=0$ indicates removing the history at location $i$ and $u_i=1$ means not intervening at $i$. 

\subsection{Expected Behavior After Intervention}
We now derive closed-form expressions for our control objectives in terms of the expected intervention-dependent intensity $\eta(t;u)$. The first objective of interest is to minimize the sum of expected rate of invasive species at our target time: $\sum_i \eta_i(T;u)$.

By the superposition theorem of point processes, the process $\mathcal{N}(t;u)$ can be decomposed into two independent point processes:
\begin{equation*}
\centering
\mathcal{N}(t;u) = \mathcal{N}_e(t;u) + \mathcal{N}_h(t;u)
\end{equation*}
$\mathcal{N}_e(t,u)$ is the counting process for events caused by the exogenous intensity from $\tau$ to $t$, and $\mathcal{N}_h(t;u)$ comprises the events generated due to the effect of previous events (history) before $\tau$. Each of these processes have associated intensities $\lambda_e(t; u)$ and $\lambda_h(t, u)$:
\begin{align}
\lambda_e(t; u) &= \mu + \underbrace{\int_\tau^t Ae^{-\omega(t-s)} \, d\mathcal{N}_e(s;u)}_{\text{from new events generated by }\mu} \\
\lambda_h(t; u) &=  \underbrace{{A} e^{-\omega (t-\tau)} (u \circ y)}_{\text{from events before }\tau}  + \underbrace{\int_\tau^t Ae^{-\omega(t-s)} \, d\mathcal{N}_h(s;u)}_{\text{from new events generated by history}}
\end{align}
Correspondingly, we have their expected values ${\eta_e(t; u) = \mathbb{E}[\lambda_e(t; u)]}$ and ${\eta_h(t; u) = \mathbb{E}[\lambda_h(t; u)]}$. For $\eta_e(t; u)$, we can write:
\begin{align}
\eta_e(t; u) &= \mu + \mathbb{E}\left[ \int_\tau^t Ae^{-\omega(t-s)} \, d\mathcal{N}_e(s;u) \right]\\
&= \mu + \int_\tau^t Ae^{-\omega(t-s)} \, \eta_e(s; u) ds
\label{eqn:exogenous-eta}
\end{align}



Using Theorem 1 from \cite{farajtabar2016multistage}, $\eta_e(t;u) = \Psi(t) \mu$ is a solution to Equation \ref{eqn:exogenous-eta} if and only if $\Psi(t) = I + \int_0^t Ae^{-\omega(t-s)} \Psi(s) ds$. For our choice of impact function:
\begin{equation}
\centering
\Psi(t) = I  + A (A-\omega I)^{-1} (e^{(A-\omega I)t}-I)
\end{equation}
Intuitively, $\Psi(t)$ is a matrix function indexed by $i,j$ which are cells. $\Psi_{i,j}(t)$ can be interpreted as the total contribution of possible invasions at cell i at time $t$ from events at j at any time before $t$ (directly and indirectly).


Additionally, according to Theorem 3 in ~\cite{farajtabar2016multistage}, by using integration by parts and the Laplace transform of point processes from~\cite{farajtabar2014shaping}, we can show that $\eta_h(t; u)= \Xi(t-\tau) A (u \circ y)$ where ${\Xi(t) = e^{(A-\omega I)t}}$.
Putting these two together we have the analytical form for our first objective:
\begin{equation}
\begin{split}
\mathbb{E}[\lambda(t; u)] = 
\Psi(t) \mu + \Xi(t-\tau) A (u \circ y)
 \end{split}
\end{equation}

For the second objective we aim to minimize the total average number of invasive species in all locations, $\sum_i \mathbb{E}[\mathcal{N}_i(T; u)]$:
\begin{equation}
\begin{split}
\mathbb{E}[\mathcal{N}_i(T; u)] = \mathbb{E}[\int_0^T d \mathcal{N}_i(s; u)] & \\
= \int_0^T \mathbb{E}[ d \mathcal{N}_i(s; u)] &= \int_0^T \eta(s; u) ds
\end{split}
\end{equation}
Therefore, if we define $\Gamma(t) = \int_0^t \Psi(s) ds$ and $\Upsilon(t) = \int_0^t \Xi(s) ds$ we have:
\begin{equation}
\begin{split}
\mathbb{E}[\mathcal{N}(t; u)] = 
\Gamma(t) \mu + \Upsilon(t-\tau) A (u \circ y).
 \end{split}
\end{equation}
It is easy to see that $\Upsilon(t) = (A-\omega I)^{-1} (e^{ (A-\omega I) t}- I)$ and $\Gamma(t) = I t + A (A-\omega I)^{-1} (\Upsilon(t) - It) $.
Intuitively, $\Gamma_{i,j}(t)$ is the cumulative invasion from $i$ to $j$ up to time $t$.

In summary we have;
\begin{align}
& \mathbb{E}[\lambda(T; u)] = \Psi(T) \mu + \Xi(T-\tau) A (u \circ y)) \label{eqn:theo-intensity-obj}\\
& \mathbb{E}[\mathcal{N}(T; u)]  = \Gamma(T) \mu + \Upsilon(T-\tau) A ( u \circ y)
\label{eqn:theo-number-obj}
\end{align}
where
%
%
\begin{align}
& \Xi(t) = e^{(A-\omega I)t} \\
& \Psi(t) = I  + A (A-\omega I)^{-1} (e^{(A-\omega I)t}-I) \\
& \Upsilon(t) = (A-\omega I)^{-1} (e^{ (A-\omega I) t}- I) \\
& \Gamma(t) = I t + A (A-\omega I)^{-1} (\Upsilon(t) - It)
\end{align}

\subsection{Optimization Formulations}
Given the closed forms we have derived for the expected behavior of the network diffusion process after intervention, we can define our first optimization problem as:
\begin{equation}
	\label{eq:optimization1}
	\begin{split}
		\underset{u}{\text{minimize}}	& \,\,\, \sum_i \Psi(T) \mu + \Xi(T-\tau) A (u \circ y)\\
    	\text{subject to:}				& \,\,\, \sum_i (1-u_i) c_i \le \mathcal{B},\\
				         				& \,\,\,\, u_i = \{0,1\} \forall i\in\{1,2,\dots,n\}
	\end{split}
\end{equation}
where $c_i$ and $\mathcal{B}$ are defined as before.

Similarly, our second objective is:
\begin{equation}
	\label{eq:optimization2}
    \begin{split}
    	\underset{u}{\text{minimize}}	& \,\,\, \Gamma(T) \mu + \Upsilon(T-\tau) A ( u \circ y)\\
        \text{subject to:}				& \,\,\,\, \sum_i (1-u_i) c_i \le \mathcal{B}, \\
        								& \,\,\,\, u_i = \{0,1\} \forall i\in\{1,2,\dots,n\}
    \end{split}
\end{equation}
The dependence on our control variable, $u$, is linear and we can incorporate effective binary optimization techniques to find the optimal intervention plan. We used the mixed integer linear programming solver offered through the \texttt{intlinprog} function in MATLAB 2016b.

\subsection{Heuristic Interventions}
Besides the optimized recommendations for intervention locations, it is also possible to choose locations on the basis of a number of heuristics. In each case, land units are considered in decreasing order of a heuristic criterion $p$, and we greedily build a set of intervention locations $U$ by adding each successive location as long as there are invasive individuals to remove there and the cost of intervening at the location can be covered with our remaining budget. This process is described in Algorithm \ref{alg:heur}, where the heuristic criterion $p$ is one of the following:

\begin{itemize}
\item \textbf{Exogenous intensity ($p_i = \mu_i$):} locations at which invasive species have the highest rate of being introduced into the network.
In invasion biology, this is known as ``propagule pressure'' and is believed to be an important component in determining whether a non-native species successfully invades a new habitat \cite{wittmann2014decomposing}.

\item \textbf{Number of events until $\tau$ ($p_i = \mathcal{N}_i(\tau)$):} locations which have seen the most number of invasions in the observation window.
Density-based eradication strategies \cite{taylor2004finding} have also been studied, especially in the context of budget availability.

\item \textbf{Intensity due to global events $\tau$ ($p_i = \lambda_i(\tau)$):} locations with the highest rate of appearance of new individuals at the intervention time.

\item \textbf{Intensity due to local events (state) ($p_i=y_i(\tau)$):} locations where there are the most actively proliferating individuals at the intervention time.
This is related to the notion of adopting an intervention strategy that balances the density and fecundity of the invasive individuals \cite{taylor2004finding}.
\end{itemize}

\begin{algorithm}
\caption{Selects intervention locations by heuristic $p$}\label{alg:heur}
\begin{algorithmic}[1]
\Procedure{HeuristicLocations}{V, $c$, $\mathcal{B}$, $\mathcal{N}(\tau)$, $p$}
\State $U \gets \emptyset$
\State $W \gets V$
\While {$\mathcal{B} > 0$ AND $W \neq \emptyset$}
	\State $u \gets \underset{i\in W}{\arg\max} \,\, p_i$
	\If {$\mathcal{N}_u(\tau) = 0$}
		\State $W \gets W \setminus u$
	\Else
		\If {$\mathcal{B} - c_u >= 0$}
    		\State $U \gets U \cup u$
    		\State $W \gets W \setminus u$
    		\State $\mathcal{B} \gets \mathcal{B} - c_u$
    	\EndIf
	\EndIf
\EndWhile
\Return $U$
\EndProcedure
\end{algorithmic}
\end{algorithm}

\section{Experiments}

\subsection{Synthetic Landscape Generation}

In order to compare the performance of different invasive species control strategies in a naturalistic setting, we generated a set of synthetic landscapes capturing different landscape structural effects on invasive spread. Each landscape consists of an $N$-by-$N$ grid of cells. The frequency of exogenous invasions at each cell $\mu_i$ is constant over time and is a uniformly distributed random variable in the range $\left[0, \mu_{max}\right]$. A small number of cells ($\theta_{\mu 1}$) have an exogenous invasion rate of $\theta_{\mu 2} \cdot \mu_{max}$ for some $\theta_{\mu 2}>1$, representing locations that act as introduction points for the invasive species.

We generate 3 different classes of landscape based on the construction of the mutual influence parameters $a_{ij}$. In all cases, $a_{ij}$ takes the form $a_{max} \cdot h_i \cdot e^{-d_{ij}^2}$. $a_{max}$ can be thought of as the establishment success rate of offspring of the invasive species given ideal conditions. However, the true establishment success may depend on the habitat suitability $h_i$ at the destination. Finally, the likelihood of offspring dispersing to a location $i$ from location $j$ is modeled as a decaying function of the squared Euclidean distance $d_{ij}$. The 3 landscape classes we generate are:
\begin{itemize}
\item \textbf{Local uniform:} Dispersal can occur only between each cell and its 8-cell neighborhood, and the habitat quality $h_i=1$ everywhere (invasives often exhibit phenotypic plasticity and can survive in a range of environments \cite{sakai2001population}).

\item \textbf{Local non-uniform:} Again, dispersal can take place only between adjacent cells, but $h_i$ varies spatially. The habitat suitability landscape matrix is generated using a mixture of $\theta_G$ 2D Gaussian functions, scaled such that $h_i\in\left[0.5, 1\right]$. Each Gaussian is characterized by $\left(\mu_x, \mu_y, \sigma_x, \sigma_y, \rho\right)$, where $\mu_x$ and $\mu_y$ are the $x,y$ coordinates of the mean, $\sigma_x$ and $\sigma_y$ are the standard deviations along each dimension, and $\rho$ is the correlation between $\sigma_x$ and $\sigma_y$.

\item \textbf{Local non-uniform with jumps:} In addition to influence between adjacent cells, a small number ($\theta_J$) of connections are allowed between non-adjacent cells, to model the effect of occasional long-range dispersal events on invasive species spread.
\end{itemize}

\begin{figure}
\centering
	\begin{subfigure}[t]{0.43\linewidth}
		\includegraphics[width=\linewidth]{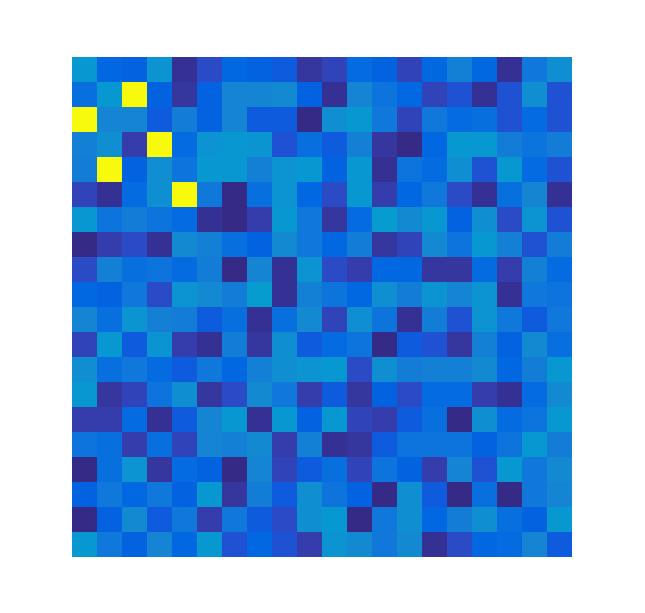}
        \caption{Exogenous intensity $\mu_i$.}
	\end{subfigure}
    ~
    \begin{subfigure}[t]{0.43\linewidth}
		\includegraphics[width=\linewidth]{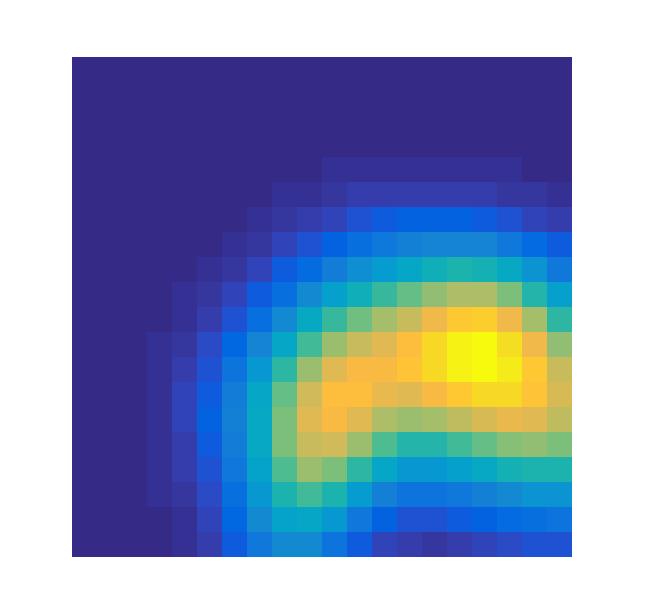}
        \caption{Habitat suitability $h_i$.}
	\end{subfigure}
\caption{Sample synthetically generated exogenous intensity matrix with 5 high invasion frequency seed points, and habitat suitability matrix generated as a mixture of 5 2D Gaussians. Brighter colors indicate higher intensity and suitability.}
\label{fig:landscape-generation}
\end{figure}

We present results for $20$x$20$ landscapes with $\mu_{max}=0.02$ with $\theta_{\mu1}=5$ invasion foci, where human-mediated introductions are responsible for on average 0.06 invasions per cell per unit time (i.e. $\theta_{\mu2}=3$). Figure \ref{fig:landscape-generation} shows a sample realization of the exogenous intensities across a synthetic landscape. For the non-uniform landscapes, we generate a habitat suitability surface using a mixture of $\theta_G=5$ Gaussians, an example of which is also pictured in Figure \ref{fig:landscape-generation}. In the local non-uniform landscape with jumps, we randomly select $\theta_J=10$ pairs of non-adjacent landscape cells between which dispersal may occur. Finally, we set the establishment success rate of the invasive species to $a_{max}=0.05$ and its death rate $\omega=0.15$.

For all landscapes, we set a finite planning horizon of $T=100$ and intervention time $\tau=50$. The intervention cost $c_i$ at each land unit is set to a fixed unit cost plus a cost proportional to the number of invasive individuals established there at time $\tau$. For each landscape type, we simulate 10 realizations of invasion cascades from $t=0$ to $\tau$. In each case, we then compute $\mathcal{B}_{tot}$ the cost of removing all the invasive individuals that have appeared in the landscape by $t=\tau$, and set the intervention budget $\mathcal{B}$ as a fixed percentage of $\mathcal{B}_{tot}$ to allow comparisons between the different realizations. In most of our problem instances, finding the optimal plan took the linear programming solver under 1 minute.

\subsection{Results}

\subsubsection{Validation of Derived Analytical Expressions}
\begin{figure}
\includegraphics[width=\linewidth]{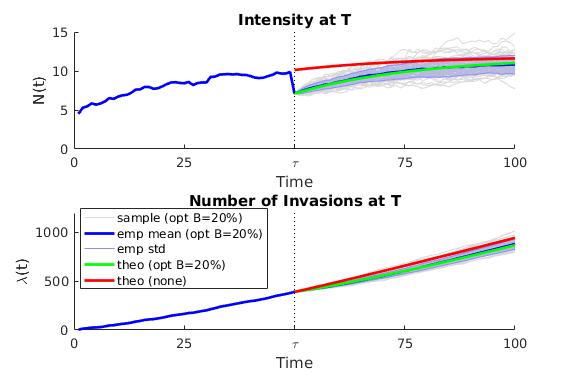}
\caption{Invasion cascades after an optimal intervention with $\mathcal{B}=0.20 \cdot \mathcal{B}_{tot}$ on a local non-uniform landscape. Simulated cascades are compared to the analytically computed intensity and number of invasions for the optimal plan.}
\label{fig:theovsemp}
\end{figure}

First, we empirically evaluate the closed-form expressions for our intervention objectives $\mathbb{E}\left[\lambda(T)\right]$ and $\mathbb{E}\left[\mathcal{N}(T)\right]$. We simulate a single realization of an invasion cascade up to time $\tau$, implement a fixed intervention $U$ and simulate many realizations of the subsequent cascade until time $T$ with which we compute the empirical intensity and number of invasions at each time $t\leq T$. We compare these to the theoretical expected intensity and number of invasions computed using Equations \ref{eqn:theo-intensity-obj} and \ref{eqn:theo-number-obj}, following the same intervention $U$. The results are shown in Fig. \ref{fig:theovsemp}. The theoretically computed values closely match the observed empirical mean values for both quantities. In the rest of our experiments, we report only the theoretically computed values.

\subsubsection{Invasives Management with a Limited Budget}

\begin{table}[h!]
\footnotesize
\centering
\begin{tabular}{|l|c|c|c|c|}
\hline
\multicolumn{5}{|c|}{\textbf{Intensity at $T$}}\\
\hline
	& \multicolumn{4}{c|}{\textbf{Budget (\%)}}\\
\textbf{Strategy} & \textbf{20}       & \textbf{40}       & \textbf{60}        & \textbf{80}        \\
\hline
none                       & 0.0$\pm$0.0   & 0.0$\pm$0.0   & 0.0$\pm$0.0   & 0.0$\pm$0.0	\\
$\mu_i$        			   & 8.6$\pm$1.1   & 16.5$\pm$1.3  & 24.3$\pm$1.4  & 32.3$\pm$2.1	\\
$\mathcal{N}_i(\tau)$	   & 8.8$\pm$1.4   & 18.0$\pm$1.6  & 27.0$\pm$2.2  & 35.6$\pm$2.6   \\
$\lambda_i(\tau)$          & 13.4$\pm$1.9  & 25.0$\pm$1.5  & 33.7$\pm$2.1  & 39.4$\pm$2.4   \\
$y_i(\tau)$		   		   & 15.7$\pm$1.2  & 27.1$\pm$1.4  & 36.5$\pm$2.4  & 40.9$\pm$2.8   \\
optimal           		   & 20.0$\pm$1.2  & 31.2$\pm$0.0  & 37.9$\pm$2.6  & 41.2$\pm$2.8	\\
all                        & 41.8$\pm$2.9  & 41.8$\pm$2.9  & 41.8$\pm$2.9  & 41.8$\pm$2.9   \\
\hline
\multicolumn{5}{|c|}{\textbf{Number of Invasions from $\tau$ to $T$}}\\
\hline
	& \multicolumn{4}{c|}{\textbf{Budget (\%)}}\\
\textbf{Strategy} & \textbf{20}       & \textbf{40}       & \textbf{60}        & \textbf{80}        \\
\hline
none                       & 0.0$\pm$0.0   & 0.0$\pm$0.0   & 0.0$\pm$0.0    & 0.0$\pm$0.0	\\
$\mu_i$					   & 11.5$\pm$1.3  & 22.4$\pm$1.4  & 33.1$\pm$1.6   & 44.2$\pm$2.2   \\
$\mathcal{N}_i(\tau)$      & 11.9$\pm$1.7  & 24.4$\pm$1.7  & 36.6$\pm$2.4   & 48.2$\pm$2.8   \\
$\lambda_i(\tau)$          & 18.0$\pm$2.3  & 33.7$\pm$1.6  & 45.6$\pm$2.2   & 53.5$\pm$2.3   \\
$y_i(\tau)$                & 21.2$\pm$1.3  & 36.7$\pm$1.3  & 49.6$\pm$2.3   & 55.6$\pm$2.8   \\
optimal           		   & 26.4$\pm$1.2  & 41.6$\pm$1.9  & 51.2$\pm$2.6   & 55.9$\pm$2.8   \\
all                        & 56.8$\pm$3.0  & 56.8$\pm$3.0  & 56.8$\pm$3.0   & 56.8$\pm$3.0   \\
\hline
\end{tabular}
\caption{Mean and SD \% reduction in the invasive species activity achieved by implementing each intervention strategy in a local uniform landscape.}
\label{table:budget-percent-reduction}
\end{table}

\begin{figure*}[h!]
\centering
\includegraphics[width=0.9\linewidth]{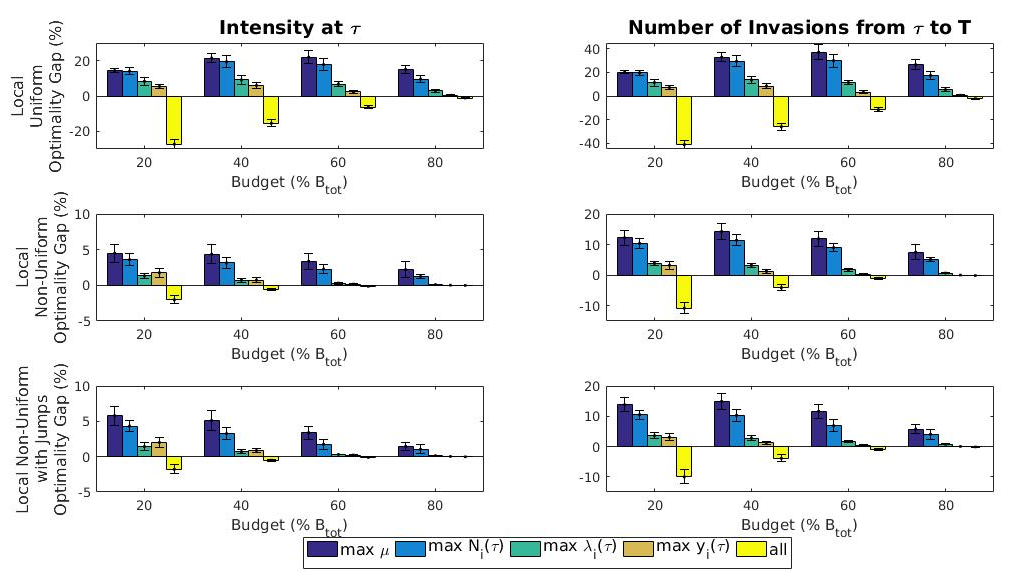}
\caption{Mean optimality gap \% for the heuristic strategies and the infeasible complete eradication intervention. Rows correspond to different landscape classes, and columns correspond to the minimization objectives.}
\label{fig:optgap}
\end{figure*}

In order to examine the impact of budgetary restrictions on the effectiveness of different invasives management strategies, we vary the intervention budget $\mathcal{B}$ available from 20\% to 80\% of $\mathcal{B}_{tot}$. For comparison, we also include the (infeasible) complete intervention in which all invasive individuals are eradicated from the landscape at time $\tau$, as well as the case in which no intervention is performed. In Table \ref{table:budget-percent-reduction} we report the \% reduction in invasives activity achieved in relation to the no-intervention case for the local uniform landscape (qualitatively similar results were obtained for the other two landscape types, and are presented in the Supplemental Information). At best, removing all invasive individuals could achieve a 41.8\% reduction in the invasion intensity at $T$ and a 56.8\% reduction in the number of invasions from $\tau$ to $T$. Correspondingly, the optimal strategy attained a 37.9\% reduction in intensity and a 51.2\% reduction in invasion events, or over 90\% of the level of control obtained by eradicating everything, with only 60\% of the budget. This demonstrates that our method has the potential to deliver significant cost savings in invasives management.

In all landscape settings and at all budget levels, the best-known feasible solution with the optimized intervention plan. Figure \ref{fig:optgap} shows the \% optimality gap of the heuristic management strategies relative to the optimal plan. The best performing heuristic approach was the selection by maximum state $y_i(\tau)$, which was consistently within 10\% of the optimal value across landscape setting and budget levels. This is in agreement with studies that have observed that the efficacy of invasive species management plans is sensitive to species life history and population growth rate (\cite{buhle2005bang}). Furthermore, we do not require a precise knowledge of the dynamics of the spread process in order to follow the $y_i(\tau)$ heuristic. Observing a trace of the invasion process or the ability to determine the life stage of observed invasive individuals may be sufficient to characterize the state of each location. This, combined with the favorable performance of the $y_i(\tau)$ heuristic, suggests it could potentially be used as a rule of thumb for planning management efforts.

\begin{figure}
\centering
    \begin{subfigure}[t]{0.28\linewidth}
		\includegraphics[width=\linewidth]{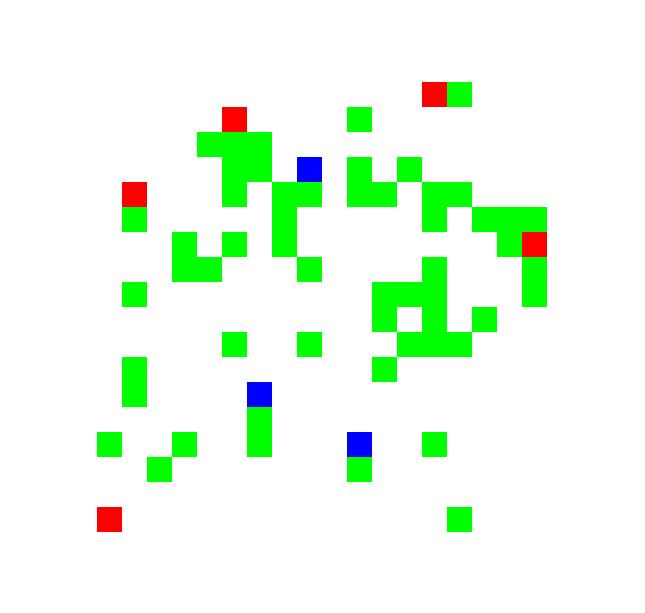}
	\end{subfigure}
    ~
    \begin{subfigure}[t]{0.28\linewidth}
		\includegraphics[width=\linewidth]{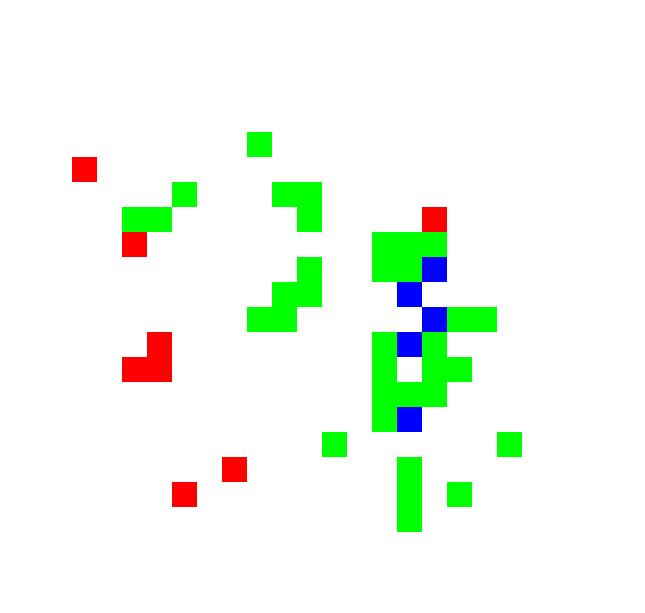}
	\end{subfigure}
    ~
    \begin{subfigure}[t]{0.28\linewidth}
		\includegraphics[width=\linewidth]{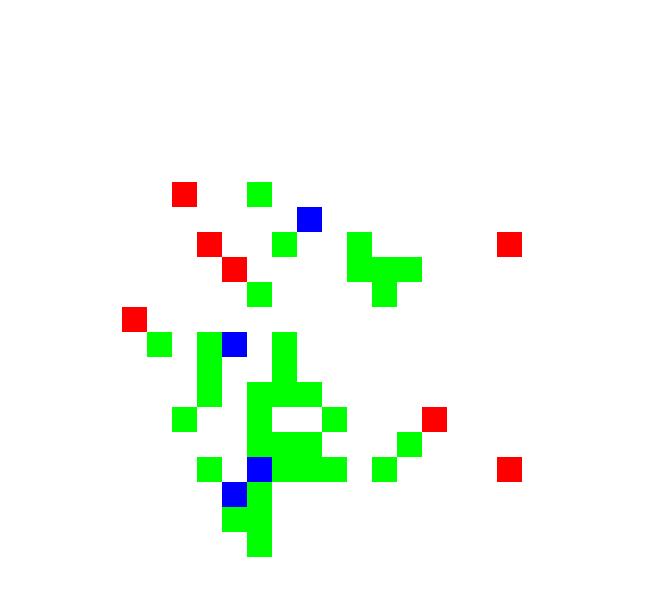}
	\end{subfigure}
\caption{Overlap between optimal intervention plans for minimizing the total invasion rate versus the total number of invasions at $T$ with $20\%\mathcal{B}_{tot}$, for (left) local uniform, (center) local non-uniform, and (right) local non-uniform + jumps landscapes. Locations minimizing intensity only (blue), number of invasions only (red) and both (green).}
\label{fig:objective-plans}
\end{figure}

Unsurprisingly, our results also show the locations recommended for minimizing each objective are different from one another (Figure \ref{fig:objective-plans}), suggesting there are possible trade-offs that may be of interest to conservation planners developing long-term strategies for invasive species management. In particular, it appears that minimizing intensity focuses intervention effort at relatively few core areas of invasion whereas minimizing the total number of invasions targets more peripheral locations.

\section{Conclusions}
We demonstrate how Hawkes processes can be used to model the dynamics of invasive species spread through a landscape. We then consider the effect of an intervention consisting of the eradication of invasive individuals at designated sites on the invasion process, which equates to history deletion in the point process. We are interested in minimizing the \emph{expected} rate of invasion and the \emph{expected} number of invasions in a finite time horizon $T$ resulting from our intervention. Our main contribution is to develop a closed-form expression for these network diffusion-related objectives after applying a given intervention plan. This introduces a novel intervention mechanism to the control of network temporal point processes, and also adds to existing methods for finding optimal intervention plans for invasive species management. Our empirical results suggest that optimized intervention plans obtained using our approach can achieve cost-effective control, and that in the absence of detailed data on the dynamics of the spread over the landscape, developing an intervention plan targeting locations with high densities of young, rapidly spreading individuals may be a good general principle.

\clearpage
\fontsize{9.5pt}{10.5pt}
\bibliographystyle{aaai} \bibliography{hawkesinvasives}

\end{document}


%
\title{Hawkes Processes for Invasive Species Modeling and Management\\Supplemental Material}
\author{Authors\\
Institution\\
}
\maketitle
\section{Additional Results}

\begin{table}[H]
\footnotesize
\centering
\begin{tabular}{|l|c|c|c|c|}
\hline
\multicolumn{5}{|c|}{\textbf{Intensity at $T$}}\\
\hline
	& \multicolumn{4}{c|}{\textbf{Budget (\%)}}\\
\textbf{Strategy} & \textbf{20}       & \textbf{40}       & \textbf{60}        & \textbf{80}        \\
\hline
none                           & 25.1$\pm$2.0  & 45.5$\pm$4.3  & 61.3$\pm$6.7  & 70.3$\pm$8.1  \\
$\mu_i$        			   & 14.4$\pm$1.1  & 21.5$\pm$2.8  & 22.0$\pm$3.9   & 15.1$\pm$2.5   \\
$\mathcal{N}_i(\tau)$	   & 14.1$\pm$1.9  & 19.3$\pm$3.3  & 17.8$\pm$3.5   & 9.6$\pm$2.1   \\
$\lambda_i(\tau)$          & 8.3$\pm$2.3   & 9.1$\pm$2.4   & 6.8$\pm$1.4    & 2.9$\pm$1.0    \\
$y_i(\tau)$		   		   & 5.4$\pm$1.3   & 6.0$\pm$1.7   & 2.3$\pm$0.9    & 0.4$\pm$0.3    \\
optimal           			   & 0.0$\pm$0.0   & 0.0$\pm$0.0   & 0.0$\pm$0.0    & 0.0$\pm$0.0    \\
all                            & -27.3$\pm$2.6 & -15.5$\pm$1.9 & -6.3$\pm$0.9   & -1.1$\pm$0.3   \\ \hline
\multicolumn{5}{|c|}{\textbf{Number of Invasions from $\tau$ to $T$}}                                                                 \\ \hline
                               & \multicolumn{4}{c|}{\textbf{Budget (\%)}}                                        \\
\textbf{Strategy} & \textbf{20}       & \textbf{40}       & \textbf{60}        & \textbf{80}        \\
\hline
none                           & 35.9$\pm$2.2  & 71.4$\pm$5.7  & 105.3$\pm$10.7 & 127.3$\pm$14.4 \\
$\mu_i$					   & 20.2$\pm$1.2  & 33.0$\pm$3.7  & 37.2$\pm$6.3   & 26.6$\pm$4.4   \\
$\mathcal{N}_i(\tau)$      & 19.7$\pm$2.3  & 29.5$\pm$4.5  & 30.1$\pm$5.6   & 17.4$\pm$3.5   \\
$\lambda_i(\tau)$          & 11.4$\pm$3.0  & 13.6$\pm$3.1  & 11.5$\pm$2.0   & 5.3$\pm$1.7    \\
$y_i(\tau)$                & 7.1$\pm$1.5  & 8.4$\pm$2.1  & 3.3$\pm$1.2   & 0.7$\pm$0.4     \\
optimal           & 0.0$\pm$0.0   & 0.0$\pm$0.0   & 0.0$\pm$0.0    & 0.0$\pm$0.0             \\
all                            & -41.4$\pm$3.2 & -26.2$\pm$2.8 & -11.7$\pm$1.7   & -2.3$\pm$0.5    \\ \hline
\end{tabular}
\caption{Mean and SD \% optimality gap of each intervention strategy in a local uniform landscape.}
\label{table:budget-optimality-gap}
\end{table}

\begin{table*}[]
\centering
\begin{tabular}{lllllllllllll}
\hline
\multicolumn{13}{c}{\textbf{Local Uniform}}\\
\hline
	& \multicolumn{2}{c}{\textbf{max $\mu_i$}}	& \multicolumn{2}{c}{\textbf{max $N_i(\tau)$}}	& \multicolumn{2}{c}{\textbf{max $\lambda_i(\tau)$}}	& \multicolumn{2}{c}{\textbf{max $y_i(\tau)$}}	& \multicolumn{2}{c}{\textbf{opt}}	&	\multicolumn{2}{c}{\textbf{all}}\\
\textbf{Budget (\%)} & \multicolumn{1}{c}{mean} & \multicolumn{1}{c}{std} & \multicolumn{1}{c}{mean} & \multicolumn{1}{c}{std} & \multicolumn{1}{c}{mean} & \multicolumn{1}{c}{std} & \multicolumn{1}{c}{mean} & \multicolumn{1}{c}{std} & \multicolumn{1}{c}{mean} & \multicolumn{1}{c}{std} & \multicolumn{1}{c}{mean} & \multicolumn{1}{c}{std} \\
\hline
20	& 8.57	& 1.10	& 8.75	& 1.48	& 13.42	& 1.95	& 15.70	& 1.18	& 20.03	& 1.24	& 41.82	& 2.93\\
40	& 16.47	& 1.25	& 17.95	& 1.56	& 25.01	& 1.48	& 27.12	& 1.36	& 31.21	& 2.03	& 41.82	& 2.93\\
60	& 24.35	& 1.43	& 26.95	& 2.16	& 33.73	& 2.14	& 36.49	& 2.36	& 37.92	& 2.60	& 41.82	& 2.93\\
80	& 32.34	& 2.10	& 35.55	& 2.56	& 39.45	& 2.41	& 40.89	& 2.77	& 41.15	& 2.84	& 41.82	& 2.93\\
\hline
\multicolumn{13}{c}{\textbf{Local Non-Uniform}}\\
\hline
	& \multicolumn{2}{c}{\textbf{max $\mu_i$}}	& \multicolumn{2}{c}{\textbf{max $N_i(\tau)$}}	& \multicolumn{2}{c}{\textbf{max $\lambda_i(\tau)$}}	& \multicolumn{2}{c}{\textbf{max $y_i(\tau)$}}	& \multicolumn{2}{c}{\textbf{opt}}	&	\multicolumn{2}{c}{\textbf{all}}\\
\textbf{Budget (\%)} & \multicolumn{1}{c}{mean} & \multicolumn{1}{c}{std} & \multicolumn{1}{c}{mean} & \multicolumn{1}{c}{std} & \multicolumn{1}{c}{mean} & \multicolumn{1}{c}{std} & \multicolumn{1}{c}{mean} & \multicolumn{1}{c}{std} & \multicolumn{1}{c}{mean} & \multicolumn{1}{c}{std} & \multicolumn{1}{c}{mean} & \multicolumn{1}{c}{std} \\
\hline
20	& 1.55	& 0.43	& 2.41	& 0.81	& 4.54	& 1.41	& 4.13	& 1.37	& 5.79	& 1.33	& 7.69	& 1.57\\
40	& 3.10	& 0.67	& 4.23	& 0.98	& 6.44	& 1.53	& 6.44	& 1.48	& 7.13	& 1.53	& 7.69	& 1.57\\
60	& 4.46	& 0.86	& 5.46	& 1.33	& 7.27	& 1.59	& 7.39	& 1.54	& 7.56	& 1.56	& 7.69	& 1.57\\
80	& 5.66	& 0.92	& 6.53	& 1.58	& 7.58	& 1.59	& 7.66	& 1.57	& 7.68	& 1.57	& 7.69	& 1.57\\
\hline
\multicolumn{13}{c}{\textbf{Local Non-Uniform with Jumps}}\\
\hline
	& \multicolumn{2}{c}{\textbf{max $\mu_i$}}	& \multicolumn{2}{c}{\textbf{max $N_i(\tau)$}}	& \multicolumn{2}{c}{\textbf{max $\lambda_i(\tau)$}}	& \multicolumn{2}{c}{\textbf{max $y_i(\tau)$}}	& \multicolumn{2}{c}{\textbf{opt}}	&	\multicolumn{2}{c}{\textbf{all}}\\
\textbf{Budget (\%)} & \multicolumn{1}{c}{mean} & \multicolumn{1}{c}{std} & \multicolumn{1}{c}{mean} & \multicolumn{1}{c}{std} & \multicolumn{1}{c}{mean} & \multicolumn{1}{c}{std} & \multicolumn{1}{c}{mean} & \multicolumn{1}{c}{std} & \multicolumn{1}{c}{mean} & \multicolumn{1}{c}{std} & \multicolumn{1}{c}{mean} & \multicolumn{1}{c}{std} \\
\hline
20	& 1.51	& 0.71	& 2.86	& 1.22	& 5.55	& 1.82	& 5.02	& 1.53	& 6.88	& 1.61	& 8.54	& 2.06\\
40	& 3.41	& 1.07	& 5.09	& 1.73	& 7.44	& 1.92	& 7.30	& 1.84	& 8.09	& 1.95	& 8.54	& 2.06\\
60	& 5.34	& 1.45	& 6.84	& 2.11	& 8.15	& 2.04	& 8.22	& 1.97	& 8.43	& 2.03	& 8.54	& 2.06\\
80	& 7.20	& 1.84	& 7.58	& 1.97	& 8.41	& 2.07	& 8.51	& 2.05	& 8.53	& 2.06	& 8.54	& 2.06\\
\hline
\end{tabular}
\caption{Percent reduction in the total invasion intensity at $T$ achieved by each strategy.}
\label{tab:percent-reduction-intensity}
\end{table*}

\begin{table*}[]
\centering
\begin{tabular}{lllllllllllll}
\hline
\multicolumn{13}{c}{\textbf{Local Uniform}}\\
\hline
       & \multicolumn{2}{c}{\textbf{max $\mu_i$}}	& \multicolumn{2}{c}{\textbf{max $N_i(\tau)$}}	& \multicolumn{2}{c}{\textbf{max $\lambda_i(\tau)$}}	& \multicolumn{2}{c}{\textbf{max $y_i(\tau)$}}	& \multicolumn{2}{c}{\textbf{opt}}	& \multicolumn{2}{c}{\textbf{all}}\\
\textbf{Budget (\%)} & \multicolumn{1}{c}{mean} & \multicolumn{1}{c}{std} & \multicolumn{1}{c}{mean} & \multicolumn{1}{c}{std} & \multicolumn{1}{c}{mean} & \multicolumn{1}{c}{std} & \multicolumn{1}{c}{mean} & \multicolumn{1}{c}{std} & \multicolumn{1}{c}{mean} & \multicolumn{1}{c}{std} & \multicolumn{1}{c}{mean} & \multicolumn{1}{c}{std} \\
\hline
20	& 11.54	& 1.32	& 11.96	& 1.67	& 18.04	& 2.31	& 21.17	& 1.32	& 26.43	& 1.19	& 56.83	& 2.96\\
40	& 22.38	& 1.39	& 24.41	& 1.66	& 33.67	& 1.61	& 36.71	& 1.30	& 41.59	& 1.93	& 56.83	& 2.96\\
60	& 33.12	& 1.62	& 36.57	& 2.39	& 45.58	& 2.16	& 49.55	& 2.32	& 51.16	& 2.57	& 56.83	& 2.96\\
80	& 44.21	& 2.15	& 48.20	& 2.77	& 53.54	& 2.32	& 55.56	& 2.76	& 55.85	& 2.85	& 56.83	& 2.96\\
\hline
\multicolumn{13}{c}{\textbf{Local Non-Uniform}}\\
\hline
       & \multicolumn{2}{c}{\textbf{max $\mu_i$}}	& \multicolumn{2}{c}{\textbf{max $N_i(\tau)$}}	& \multicolumn{2}{c}{\textbf{max $\lambda_i(\tau)$}}	& \multicolumn{2}{c}{\textbf{max $y_i(\tau)$}}	& \multicolumn{2}{c}{\textbf{opt}}	& \multicolumn{2}{c}{\textbf{all}}\\
\textbf{Budget (\%)} & \multicolumn{1}{c}{mean} & \multicolumn{1}{c}{std} & \multicolumn{1}{c}{mean} & \multicolumn{1}{c}{std} & \multicolumn{1}{c}{mean} & \multicolumn{1}{c}{std} & \multicolumn{1}{c}{mean} & \multicolumn{1}{c}{std} & \multicolumn{1}{c}{mean} & \multicolumn{1}{c}{std} & \multicolumn{1}{c}{mean} & \multicolumn{1}{c}{std} \\
\hline
20	& 5.10	& 1.06	& 6.65	& 1.35	& 12.11	& 2.06	& 12.63	& 2.22	& 15.40	& 1.79	& 24.53	& 2.38\\
40	& 10.09	& 1.49	& 12.34	& 1.60	& 18.77	& 2.12	& 20.30	& 2.14	& 21.28	& 2.16	& 24.53	& 2.38\\
60	& 14.65	& 1.96	& 16.85	& 2.09	& 22.35	& 2.22	& 23.49	& 2.22	& 23.70	& 2.27	& 24.53	& 2.38\\
80	& 18.80	& 1.60	& 20.55	& 2.53	& 23.92	& 2.36	& 24.41	& 2.35	& 24.43	& 2.35	& 24.53	& 2.38\\
\hline
\multicolumn{13}{c}{\textbf{Local Non-Uniform with Jumps}}\\
\hline
       & \multicolumn{2}{c}{\textbf{max $\mu_i$}}	& \multicolumn{2}{c}{\textbf{max $N_i(\tau)$}}	& \multicolumn{2}{c}{\textbf{max $\lambda_i(\tau)$}}	& \multicolumn{2}{c}{\textbf{max $y_i(\tau)$}}	& \multicolumn{2}{c}{\textbf{opt}}	& \multicolumn{2}{c}{\textbf{all}}\\
\textbf{Budget (\%)} & \multicolumn{1}{c}{mean} & \multicolumn{1}{c}{std} & \multicolumn{1}{c}{mean} & \multicolumn{1}{c}{std} & \multicolumn{1}{c}{mean} & \multicolumn{1}{c}{std} & \multicolumn{1}{c}{mean} & \multicolumn{1}{c}{std} & \multicolumn{1}{c}{mean} & \multicolumn{1}{c}{std} & \multicolumn{1}{c}{mean} & \multicolumn{1}{c}{std} \\
\hline
20	& 4.30	& 1.43	& 7.02	& 1.95	& 12.84	& 2.49	& 13.26	& 2.11	& 15.86	& 2.32	& 24.25	& 3.81\\
40	& 9.54	& 2.03	& 13.11	& 2.92	& 19.01	& 2.95	& 20.18	& 2.98	& 21.22	& 3.14	& 24.25	& 3.81\\
60	& 14.59	& 2.53	& 18.22	& 3.59	& 22.24	& 3.58	& 23.19	& 3.54	& 23.47	& 3.62	& 24.25	& 3.81\\
80	& 19.79	& 3.22	& 21.11	& 3.51	& 23.57	& 3.80	& 24.12	& 3.78	& 24.15	& 3.79	& 24.25	& 3.81\\
\hline
\end{tabular}
\caption{Percent reduction in number of invasions between $\tau$ and $T$ achieved by each strategy.}
\label{tab:percent-reduction-number}
\end{table*}